# Coupled effects of applied load and surface structure on the viscous forces during peeling

Charles Dhong and Joëlle Fréchette*

Department of Chemical and Biomolecular Engineering, Johns Hopkins University, Baltimore, Maryland 21218, United States

* Corresponding author. Telephone: (410) 516-0113. Fax: (410) 516-5510. E-mail: jfrechette@jhu.edu.

**Abstract**

Tree frogs are able take advantage of an interconnected network of epithelial cells in their toe pads to modulate their adhesion to surfaces under dry, wet, and flooded environments. It has been hypothesized that these interconnected drainage channels reduce the hydrodynamic repulsion to facilitate contact under a completely submerged environment (flooded conditions). Using a custom-built apparatus we investigate the interplay between surface structure and loading conditions on the peeling force. By combining a normal approach and detachment by peeling we can isolate the effects of surface structure from the loading conditions. We investigate three surfaces: two rigid structured surfaces that consist of arrays of cylindrical posts and a flat surface as a control. We observe three regimes in the work required to separate the structured surface that depend on the fluid film thickness prior to pull out. These three regimes are based on hydrodynamics and our experimental results agree with a simple scaling argument that relates the surface features to the different regimes observed. Overall we find that the work of separation of a structured surface is always less than or equal to the one for a smooth surface when considering purely viscous contributions.

1. Introduction

Throughout nature, animals have taken advantage of structured surfaces to mediate their adhesion in dry, wet, and flooded environments. Tree frogs are an interesting case because they display good adhesion in all these conditions. In a completely flooded environment tree frogs have been shown to exhibit strong control over their locomotion, for example being able to move without slipping without interlocking supports.[1-5] They have also been shown to adhere upside-down under flowing (5 mL/min to 4L/min) water.[2] A key feature that enables this control is suspected to be the structured hexagonal array of epithelial cells on their toe pads. The epithelial cells have a soft (5-15 MPa), keratinized outer layer[6] and form pillars on the toe pads that are approximately 10 μm in diameter, 10 μm deep and are separated by 1 μm, and mucus is secreted through these channels to enhance adhesion.[7, 8] The role of the channels in completely flooded environment, however, is not well-understood. The spacing between the epithelial cells creates an interconnected network of channels that has been hypothesized to aid in the removal of fluid from the toe pad, reducing the hydrodynamic repulsion and as a consequence reduce the time necessary to make contact.[1, 4] The interplay between surface structure and hydrodynamic interactions has implications in several fields[9] beyond the understanding of tree frog adhesion. For example, in the design of structured surfaces for drag reduction in underwater propulsion[10], for the flow and solute transport through cracks in hydrofracturing[11], in the design of tire treads to prevent hydroplaning[12, 13], or to minimize viscous losses in micro- and nanoscale resonators that operate in fluid environments[14, 15]. Micro- and nanoscale roughness or structure also dictate slip at the solid-liquid interface.[16, 17]

To understand how topography can modulate the force to separate surfaces in flooded environments, it is important to consider a peeling motion during detachment since many animals, including tree frogs, detach in a peeling mode.[18] During peeling, the detachment occurs by gradually increasing the peeling angle as a crack propagates across the toe pad. Structured surfaces offer multiple advantages when trying to modulate peeling forces, they can blunt the crack front[19] or force the arrest of crack propagation at feature boundaries[20]. Detachment via peeling is also desirable because the adhesion force can be modulated by varying the peel angle.[21-24] Geckos[25] and tree frogs[3] have both been shown to splay their limbs to control their adhesion to surfaces where the peel angle is kept low to maintain contact and increased to pull out. In biomimetic systems, the effect of structured surfaces on peeling has been well-studied in dry[26-28] and wet[28, 29] environments, but less so in completely flooded conditions.

The coupling between approach and pull out is another important characteristic that needs to be considered to understand detachment between surfaces under flooded conditions. In contrast to the adhesion force measured in air where there is negligible work necessary to make contact, viscous forces are present when submerged in fluid and affect both approach and pull out, especially if it is necessary to make contact rapidly. For example, significant viscous drag can prevent surfaces from reaching contact quickly or would require a significant applied load.[30] The viscous force required to move two flat surfaces in a fluid is described by the Stefan equation (Eqn 1)[31], which can be derived from Reynolds' theory[32] and has been solved for several plate geometries[30], including in the presence of misalignment such as tilt[33]. In the context of work required to separate surfaces in flooded environments, the initial loading sets the fluid film thickness prior to pull out, and as a consequence influences whether or not conservative contact forces are present and the magnitude of the viscous forces during pull out. Therefore, when investigating the role of surface structure on the work required to separate surfaces, it is important to control for the loading conditions to decouple its effect from that of the surface structure.

Under flooded conditions, the presence of surface structure can have profound effects on the work and time required to make contact. For instance, the spacing between surface features can facilitate drainage and reduce the repulsive hydrodynamic forces present when two surfaces are brought together in a viscous fluid.[34,35] A scaling argument, initially proposed by Persson[34], relates the decrease in time to contact to the feature sizes via a single length scale ($h_o$), a parameter that captures the key dimensions of the surface features. A similar argument can be reached from an effective permittivity and Darcy's law using a porous media analysis.[36-38] According to this limiting scaling argument (see Section 2), if the fluid film thickness is larger than $h_o$ the fluid drainage is radial, there is no flow through the surface features, and the viscous forces are the same as the ones for flat surfaces. For fluid film thickness smaller than $h_o$, the fluid drainage goes through the surface features and the viscous forces are lower than the ones for smooth surfaces. We previously verified experimentally this scaling argument for the cross-cylinder geometry via the direct measurement of the hydrodynamic force between a flat and a structured surfaces in the surface forces apparatus (SFA).[35]

There are multiple reports on the role of surface structure on the detachment forces in viscous fluids. An enhancement in the detachment force was observed when peeling a surface patterned with an array of PDMS posts in a viscoelastic fluid.[29] It was suggested that the origin of the enhancement in adhesion energy was due either to viscous dissipation, crack blunting, or anchoring of the fluid to the structure. Measurement by Drotlef et al.[39] of the adhesion and friction forces of elastic microstructured surfaces in fluids showed that the presence of surface features increased the friction forces, possibly due to boundary contact being facilitated by drainage through the surface structure. They also observed that an increase in fluid viscosity had no effect on the force and concluded that there were no viscous contributions to the peak in the adhesion force. In the case of oil capillary bridges on top of micropillars, it has been predicted that for pillars larger than 10 μm the major contribution to the adhesive force should be both viscous and the contribution from the Laplace pressure in the liquid bridge.[40] There have been many instances when investigating structured surfaces where viscous contributions have been considered[29, 39-41] but there is a need to isolate the viscous contributions from other effects such as capillary or van der Waals interactions, and elasticity of the surfaces[42]. Moreover, one common features of the previous reports is that the detachment force is tested after the sample has been brought into contact with a substrate without control for the loading conditions or the time necessary to make contact. The loading conditions (viscosity, applied load, and loading time) dictate the fluid film thickness prior to pull out and, in turn, the work required to separate the surfaces.

Here we detail our investigation of how the interplay between surface structure and loading conditions affect the viscous contribution in a peeling mode. We use rigid structured surfaces in Newtonian fluids and control the fluid film thickness prior to pull out by varying the viscosity, applied load, and loading time during approach. By following this protocol we can compare the work of separation for identical loading conditions to isolate the effect of surface structure. We observe that the presence of surface features facilitate contact and decrease the work of separation. We discuss our results in the context of the scaling of the lubrication approximation for structured surfaces.

**2. Effect of surface structure on the fluid film thickness during approach.**

Consider the approach between a surface with a periodic array of pillars and a smooth wall in the lubrication limit, where the fluid film thickness can be described by the Reynolds equation. We assume that the surfaces are rigid, that inertial effects are negligible (Re < 1), and that the plate area is much larger than the fluid film thickness. We follow the analysis of Persson[34] and hypothesize the presence of three different limiting regimes for fluid flow, illustrated in Figure 1. First, at large separations (short times) there is a far-field regime where the fluid flows radially and not through the structure. Second, as the separation decreases further the pressure in the gap increases and becomes sufficiently large to favor fluid flow through the structure instead of radially. As a consequence, preferential drainage of fluid through the structure yields smaller fluid film thickness for a given loading time than predictions based on smooth surfaces. Finally at small separations the hydrodynamic interactions with individual pillars dominate, and we recover the Reynolds equation but for an array of individual pillars (effective lower surface area). We denote this final stage the near-field regime. Therefore, as a fluid film thickness decreases during approach there should be a transition between radial flow and flow through the surface features. In the limit where the pillar height is much greater than the channel width (D>>W in Figure 3) the thickness of the fluid film thickness necessary for this transition can be estimated as $h_o = W\left(\frac{D}{W+d}\right)^{1/3}$, see Table 1. We previously characterized this transition[35, 43] and its relationship with $h_o$ for the hydrodynamic force present in the approach between a surface with a hexagonal array of cylindrical posts and a smooth surface. The experiments were performed in the Surface Forces Apparatus with curved surfaces in the cross-cylinder geometry.

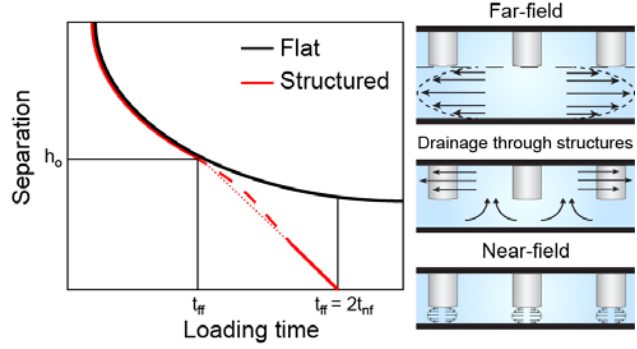

**Figure 1.** Schematic of the change in separation with loading time for a smooth and structured surface. The dashed line represents an interpolation between two regimes.

The scaling arguments derived in Ref 34 relates the characteristic times for the different flow regimes illustrated in Figure 1 to the surface features and loading conditions. First in the far-field regime the fluid flow conditions are the same as for a smooth surface with the fluid film thickness determined from the top of the posts. The change in the fluid film thickness is then obtained from the Stefan equation (Eqn 1). The Stefan equation describes the instantaneous velocity (dh/dt) when two flat plates of area A separated by a fluid film of thickness h are brought closer under an applied load ($F_N$) in a fluid of viscosity $\mu$.[31, 34] We hypothesize that the change in separation can be described by Eqn 1 until the film thickness reaches $h_o$ at the end of the far-field regime (at t = $t_{ff}$). We obtain $t_{ff}$ by solving Eqn 1 for the case of a constant $F_N$ and plates that are initially very far apart to obtain a relationship between the loading time, t, and the instantaneous surface separation, fluid film thickness h, and then set the separation to $h_o$ to find the limit of the far-field regime ($t_{ff}$), see Eqn 2.

$$\frac{dh}{dt} = -\frac{F_N}{2\mu A} h^3 \qquad (1)$$

$$t_{ff} \cong \mu \frac{A^2}{F_N} \frac{1}{h_o^2} \qquad (2)$$

We can predict the onset of the near-field regime ($t_{nf}$), with the assumption that once $h_o$ is reached the fluid flow is only through the structure and independent on the fluid film thickness. By using Eqn 1 with a constant h = $h_o$ at all times on the right hand side and with the boundary condition that at t=$t_{ff}$, h=$h_o$ we can find the time necessary to reach boundary contact (in the limit where t=$t_{nf}$ at h=0), given by Eqn 3. However, more realistically in the near-field regime the hydrodynamic interactions between individual posts and the surface would dominate at small separation and lead to dh/dt to decrease asymptotically as the fluid film thickness decreases, preventing boundary contact.

$$t_{nf} \cong \mu \frac{A^2}{F_N} \frac{1}{h_o^2} + t_{ff} = 2t_{ff} \tag{3}$$

Therefore based on Eqns 2-3 we would predict that 1) the limit of the far-field regime is inversely proportional to $h_o^2$, and 2) the time to reach boundary contact ($t_{nf}$) should be twice the time spent in the far-field regime for any surface structure.

### 3. Experimental Details.

**Sample Preparation.** All the samples investigated consist of 20 μm of SU-8 2007 (MicroChem) supported by a glass coverslip (Schott D263M, 22x22 mm, 0.13-0.16 mm thickness). The structured surfaces have features only on the top 10 μm that consist of cylindrical pillars in a hexagonal array (see Figure 3 and Table 1). For all the samples the final SU-8 thickness is achieved in two sequential 10 μm coatings on coverslips. Traditional microfabrication techniques are used to pattern the surface features. First, square glass coverslips are cleaned using an isopropyl alcohol/ethanol rinse followed by a dehydration bake at 200° C for 10 minutes. Then a layer of SU-8 2007 is spin coated at 1700 rpm for one minute to produce a 10 μm thick layer. The square substrate requires manual edge bead removal with a razor blade, which is followed by a pre-exposure bake on a hot plate at 95° C for 3 minutes and then exposure to a UV light at 140 mJ/cm². The base layer requires no mask or developing and the sample is hard-baked at 200° C for 10 minutes immediately after exposure. After cooling, a second layer of SU-8 is deposited using the same steps as the initial layer, but a chrome mask is used during the exposure step to create the surface patterns and a simple transparency mask is used for the smooth surface. For all the samples, the feature area is 14 mm x 14 mm and thus does not cover the entire coverslip substrate surface. Effort was made to manually center the mask with the coverslip, but there is sample-to-sample variation of order 1 mm from the edge of the patterned region to the edge of the coverslip. UV exposure is followed by a post-exposure bake at 95° C for 5 minutes and then immersion in SU-8 developer for 3 minutes with gentle manual agitation. Samples are then rinsed in isopropyl alcohol and hard baked at 200° C for 10 minutes. Pattern formation and layer thicknesses are verified using confocal imaging and profilometry. The bottom surface in the peeling experiments consists of a glass coverslip onto which a thin fluoropolymer layer of 1.55% CyTop (Bellex International Corporation) is spin coated at 5000 rpm for one minute and then annealed in an oven for 15 minutes at 180°C.

**Materials.** The fluids in the bath are Newtonian silicone oils (PMX-200, Xiameter). Two viscosities are investigated, 200 cSt (0.965 g/mL) and 1000 cSt (0.968 g/mL). The silicone oils were used as received. The combination of rigid SU-8 as the surfaces and silicone oils as the fluid, along with working under completely flooded conditions allow us to neglect other type of interactions such as van der Waals, electrostatic, capillary. The Hamaker constant a SU-8 – Silicone Oil – CyTop system is negligible, see ESI for estimates of the Hamaker constant and interfacial energy.

4. **Peeling Apparatus.**

A custom-built peeling apparatus, illustrated in Figure 2, was designed to measure the force required to peel the samples in a completely flooded environment for different loading conditions. The apparatus is based on the designs of Ghatak et al.[27]. The experiments are performed in two distinct, but continuous, phases inside a bath filled with fluid. An overview of the key features of the apparatus, and experimental protocol is described here, with additional details about the setup and the load cell available in the ESI.

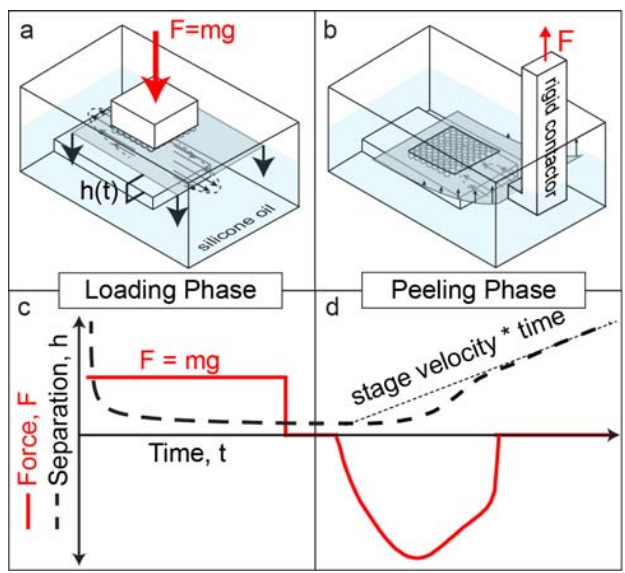

**Figure 2.** Schematic of the peeling apparatus. (a) Loading phase where a normal load is applied to decrease the fluid film thickness between the sample and bottom substrate. (b) Peeling phase where one side of the sample is moved upward by a rigid contactor driven by a motorized stage and mounted onto a load cell. (c) Illustration of the change in the fluid film thickness (separation) due to the applied load as a function of time during the loading phase. (d) Illustration of the change in separation (fluid film thickness) and measured force during the peeling process. The difference in the contactor velocity and the motor velocity gives rise to a force measured by the load cell. The fluid film thickness (separation) in (d) varies spatially and is largest on the side near the contactor.

**Loading phase.** First, in the loading phase the surfaces are initially far apart (approx. ~500 μm) and a fixed mass (0.05kg or 0.208kg) is applied on the top surface as a weight for a set amount of time (Fig. 1a,c). The constant load brings the top surface closer to the bottom surface and sets the fluid film thickness prior to the peeling phase. Therefore we control the sample-to-substrate separation prior to pull-out by changing the viscosity, the applied load, and the loading time. In the current experiments the separation between the sample and the bottom surface cannot be measured directly. There is also always a small tilt present between the two surfaces, even after careful effort to align the surfaces. A small tilt can significantly decrease the fluid film thickness from that predicted for a smooth falling plate in the lubrication limit.[30, 33] Therefore we report the loading conditions (viscosity, loading time, and applied load) as an indication for the fluid film thickness prior to measurement of pull out forces rather than absolute values of film thickness. The weight is lifted within three seconds after the end of the loading phase and the peeling measurements are performed (Figure 2b,d).

**Peeling phase.** Once the weight is pulled away from the back of the coverslip, the peeling phase starts. In the peeling phase, the flexible coverslip (flexural rigidity = 0.03 Nm) is peeled off the substrate by a rigid contactor moving at a constant drive velocity of 300 µm/s and connected to a load cell. As viscous forces scale with the velocity, we selected 300 µm/s as the drive velocity to exploit the full range of the load cell while remaining in the lubrication regime. While the drive velocity is constant, the actual velocity at which the sample and substrate separates is less than the drive velocity because of the hydrodynamic drag, and varies both with time and position. As an upper bound, we estimate the Reynolds number to be Re <1 based on a film thickness of order microns, a length of 14 mm, and a peeling velocity less than 1mm/s. Throughout the peeling process the bending of the coverslip remains in the small angle limit (<5°). In this limit we do not have to consider the potential energy from the movement of an inextensible film with an applied force.[45] Additional possible contributions to the forces measured are the elasticity of the coverslip during bending, the viscous forces from the fluid film, the compliance of the polymer film, and the conservative surface forces (such as van der Waals interactions)[29, 45-47]. In our system there is negligible contribution from conservative surface forces (e.g. electrostatic or van der Waals interactions), and the SU-8 polymer film employed here is non-compliant (E=5.6 GPa)[48]. If we consider the two remaining contributions: the elasticity of the coverslip and the viscous contributions, we find that by using a rigid backing[49] the elastic work term is much smaller than the viscous work[45]. An estimate of the bending contribution is ≪1% of the entire work (since the elastic stress term is usually much smaller than the Young's modulus) and is thus negligible.[27, 29]

5. **Results and Discussion**

To isolate the role played by the surface features on the peeling forces we aim to eliminate any non-viscous contributions to the forces measured, including deformation of the pillars. In the context of tree frogs adhesion, however, deformation of the epithelial cells could also play an important role in modulating adhesion.[34] We conducted the peeling measurements in fluids of two viscosities, two masses acting as weights for a range of loading time that vary between 5-800 s. We therefore explore a range of loading conditions of viscosity/(mass ∗ loading time) that spans over three orders of magnitude (from $2 \times 10^{-2}$ to $6 \times 10^{-6}$ cSt/kg*s). This quantity is proportional to the square of the predicted film thickness for a flat surface. Since we expect the onset of different regime behaviors to depend on separation set by the substrate, we pick a maximum range of loads that can reliably be supported by our apparatus. The magnitude of the hydrodynamic forces will depend on the velocity, but based on our prior work[35] we would not expect the velocity to change the alter the contribution of the surface structure to the drag force in a low Reynolds number regime.

Three sets of samples were fabricated, the first two have surface features consisting of a hexagonal array of cylindrical pillars and the third is a flat surface that acts as a reference and control (see Fig. 3). The two structures investigated are identical in all dimensions except for their channel width (3µm and 10µm), the dimensions of the surface features are listed in Table 1. Also listed in Table 1 is $h_o$, which represents predictions for the fluid film thickness at the transition between radial fluid and drainage through the structure (see Section 2). This parameter is derived from a scaling argument in the lubrication limit (inherently 2D), and as such assumes a limiting geometry in the remaining dimension[34]. Our structures are not in such limiting geometries so we give a range of $h_o$ in Table 1 – one determined by the width of the channels and one by the diameter of the posts.

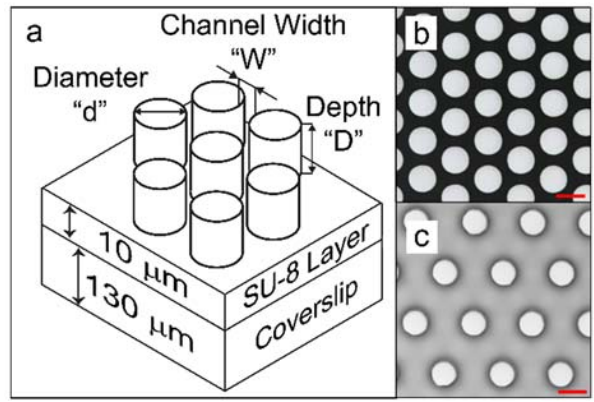

**Figure 3.** (a) Schematic and (b,c) optical microscopy images of the structured surfaces investigated. The scale bar corresponds to 10 µm. In (b) d=D=10 µm, and W=3 µm, in (c) W=D=d=10 µm.

**Table 1.** Feature sizes and estimates of $h_o$ for the samples investigated.

| Sample | Thickness (μm) | Diameter, d (μm) | Depth, D (μm) | Width, W (μm) | $h_o = W\left(\dfrac{D}{W+d}\right)^{1/3}$ (μm) | $h_o = D\left(\dfrac{W}{W+d}\right)^{1/3}$ (μm) |
|---|---|---|---|---|---|---|
| Flat | 20 | N/A | N/A | N/A | N/A | N/A |
| W=3 μm | 20 | 10 | 10 | 3.0 | 2.8 | 6.1 |
| W=10 μm | 20 | 10 | 10 | 10 | 8.0 | 7.9 |

**Force Curves.** Representative force curves for the range of loading conditions investigated here are shown in Figure 4. We see that for all loading conditions and samples, the measured forces display the same qualitative features: the forces rise rapidly, reach a peak, and drop abruptly. These general features are qualitatively similar to the ones observed for other peeling[29, 50] and normal force measurements[44, 51] in viscous fluids with smooth surfaces. In the force curves, the position and magnitude of the peak forces are instrument specific and characteristic of the load cell employed. The effect of a compliant load cell on the measurements of a viscous force has been studied in multiple systems before, including probe tack measurements[51] and the surface force apparatus[44].

Introducing structured surfaces tends to decrease the magnitude of the force measured. This observation is in sharp contrast to previous work reported in the literature for the peeling force in the presence of a structured surface in fluid environments. It has been suggested that surface features could enhance the peeling force through increased viscous dissipation or by "anchoring" liquid bridges (pinning).[29] Our experiments with rigid structured surfaces rule out the viscous flow hypothesis and suggest that either anchoring of the fluid or elasticity of the structured surface is necessary for the enhancement of peeling force. For the portion of the force curves past the force peak, we observe an abrupt decrease in the measured force. This portion of the force curve is attributed to crack propagation (see ESI). For soft patterned surfaces it has been observed that the force decreases and then reaches a plateau, which has been attributed to cavitation[52] or fingering instability (Saffman-Taylor[29] or elastic[53]).[54] Here we do not clearly see such features because no interface with a different viscosity is present in a completely flooded environment (no Saffman-Taylor type instability) and the high modulus of the SU-8 also hinders the formation of elastic instabilities. We also do not observe features that are consistent with cavitation because either the stress on the fluid film is not sufficient to induce cavitation or we are unable to resolve cavitation in our force curves. Finally, we do not observe a characteristic sequence of peaks that is typical for a series of crack arrest and propagation events caused by the surface features that have been observed in the absence of fluid.[27, 55-57] In most cases investigated here the two surfaces are not in contact; second we are working in a non-adhesive system (SU-8/silicone oil/fluoropolymer) and the stress decay length in fluids would be essentially zero and these peaks are visible when the spacing of patterns is of order the stress decay length[57]. Our observations are in agreement with the work of Patil et al.[29], where no sequence of crack arrest peaks are observed in a similar fluid experimental system.

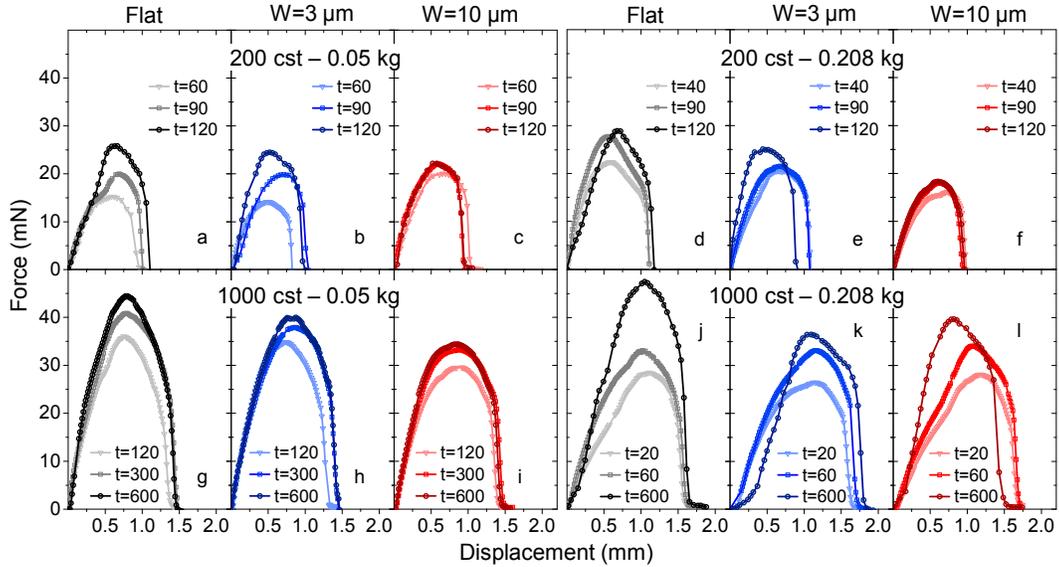

**Figure 4.** Representative force curves for the different loading conditions investigated. The structures investigated are the same for each columns. Time represents the loading time (in seconds) from the loading phase. The displacement refers to the motor displacement during the peeling phase.

Several features in the force curves, such as the effect of the load, loading time and viscosity, are self-consistent and in agreement from predictions in the lubrication limit. In all samples, the magnitude of the force increases with increasing applied load and loading time, except for the W=10μm surface in 200 cSt - 0.208 kg (Fig. 4f), where increasing the loading time has a very small effect on the force curves. The intial increase in the force before the peak is reached has been shown to depend both on the rigidity of the system and the moment arm from the rigid contactor to the structures.[26, 58] The rigidity of all our samples is the same, but the structures have slight (~ 1 mm ≅ <5%) variation in positioning from the edge of the coverslip. This misalignment, combined with slight variations in apparatus placement of the rigid contactor, alters the moment arm and leads to sample-to-sample variations in the force curves. For example, in Figure 4g-i, the force curves are all from a single sample within an individual panel, illustrating the similarity in the force profile when the alignment is the same. In contrast, the force curves in Fig. 4j-l, come from two different samples, illustrating the effect of different moment arms and sample-to-sample variations.

**Work of Separation.** For all the force curves we integrate the force versus motor displacement to obtain a work of separation for the two interacting surfaces (Figure 5). Studying the work of separation is convenient because 1) in contrast to the force curves it is not instrument dependent, the work is unaffected by the compliance of the load cell[44], 2) it captures the viscous forces for the whole separation process and not only the initiation or propagation of a crack, and finally 3) it is unaffected by small differences in the moment arm. We want to distinguish this work of separation from the work of adhesion, the latter being a thermodynamic quantity based on conservative forces, whereas the viscous forces investigated here are dissipative.

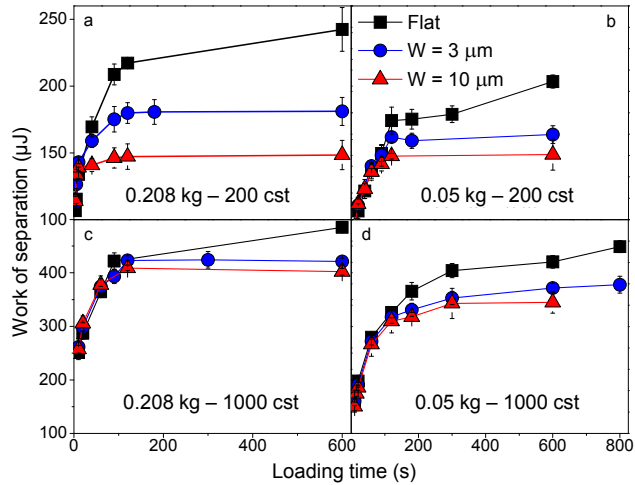

**Figure 5.** Work of separation as a function of loading times. Each panel corresponds to a different combination of viscosity and applied load. Each data point represents at least three different samples tested in triplicate. Note that the scale of the y-axis is different for the two viscosities investigated.

In Figure 5 the work of separation is plotted as a function of the loading time. For an individual panel an increase in loading time should lead to a decrease in the fluid film thickness prior to pull out. We observe that increasing the viscosity, for the same applied load and loading time, leads to an increase in the work of separation. Two factors have to be taken into consideration to explain this observation: 1) the effect of viscosity in setting the fluid film thickness prior to peeling, and 2) the differences between normal and peeling motion. First, if we only consider normal motion for both approach and retraction, there should be no effect of viscosity on the work of separation for the same loading conditions. This is because the viscosity dependence of both the drag force that sets the fluid film thickness prior to pull out and for the drag force during retraction cancels out for normal motion. The same is expected if we take into account the compliance of the load cell and model our system as a spring (the load cell) in series with a dashpot (the hydrodynamic force of the interacting surfaces).[44] In the case of peeling, previous theoretical work[49,59,60] predicted that the drag force depends on $\mu^{1/4}$ compared to being proportional to $\mu$ for normal motion. This weaker dependence of the drag force on viscosity when going from a normal to a peeling motion can therefore explain the increase in the work of separation with viscosity for the same applied load and loading time. The approach-detachment cycle is no longer reversible due to the difference in the mode of motion leading to the lower work of separation required to separate surfaces via peeling.

For a given fluid viscosity and applied load, the work of separation initially increases rapidly with loading time and then slows down or even reaches a plateau at long times. In the case of the flat surface a plateau in the work of separation is not typically observed (see Fig. 5d). In contrast, for the two structured samples, the work of separation reaches a plateau. For W=10 μm, the plateau is first observed at shorter loading times than for the W=3 μm surface. We also see that for each panel, the work of separation at long times decreases when going from a flat surface to the W=3 μm surface, and then to the W=10 μm surface. Another clear feature is that for a given viscosity and applied load the work of separation at short loading times is the same for the three surfaces investigated.

We describe the dependence of the work of separation on loading time for the structured surfaces based on the two characteristic times introduced in section 2 ($t_{ff}$ and $t_{nf}$) and illustrated in Figure 1. First, for a given viscosity and applied load we find the longest loading time for which a structured surface has the same work of separation as the flat surface, which we denote $t_{ff}$, the limit of the far-field regime. Second, for a given structured surface we find the loading time at which the plateau in the work of separation is first observed and denote it $t_{nf}$, the onset of the near-field regime. The values for $t_{ff}$ and $t_{nf}$ are listed in Table 2 and were determined for each panel in Figure 5. For a quantitative determination of the two characteristic times, we employed the Wilcoxon rank sum method[61] (threshold of P=0.05 in all cases except P=.06 for $t_{ff}$ of W= 10 μm 1000 cst 0.05 kg). To determine $t_{ff}$ the method was employed to find significant differences between the work of separation between a flat and a structured surface. To determine the onset of the plateau region, i.e. $t_{nf}$, the method was employed to find the loading time at which an increase in loading time no longer leads to an increase in the work of separation.

**Table 2.** Values for $t_{ff}$ and $t_{nf}$ for the two structures.

| Loading conditions | | W=10 μm | | W=3 μm | |
|---|---|---|---|---|---|
| viscosity | mass | $t_{ff}$ (s) | $t_{nf}$ (s) | $t_{ff}$ (s) | $t_{nf}$ (s) |
| 1000 cSt | 0.05 kg | 120 | 300 | 180 | 300 |
| | 0.208 kg | 60 | 120 | 60 | 120 |
| 200 cSt | 0.05 kg | 60 | 90 | 90 | 120 |
| | 0.208 kg | 10 | 40 | 40 | 90 |

The presence of a plateau for structured surfaces at long loading times in Figures 5-6 suggests that boundary contact is reached in the near-field regime. In contrast to the structured surfaces, we did not observe a plateau in the work of separation as the loading time is increased for flat surfaces. For smooth surfaces and considering only hydrodynamics during approach, dh/dt asymptotically decreases as h decreases. Therefore, boundary contact should not be reached and the work of separation should keep increasing with loading time, which is what we observe for flat surfaces. It is found, however, that surface roughness[62], or certain geometries[63] are often sufficient for the fluid between two surfaces to squeeze-out and the surfaces to reach contact[62, 64, 65, 66] in a finite amount of time. In the absence of surface deformation, if surfaces reach boundary contact, any longer loading should not change the work of separation. Therefore based on the fact that we observe a plateau in the work of separation we suspect that boundary contact is achieved at (or near) $t_{nf}$ for the structured surfaces.

We aim to relate the characteristic loading times to the feature dimensions reported in Table 1 and to the flow regimes outlined in Section 2. We assign the characteristic times in Table 2 to the loading time spent in the far-field regime ($t_{ff}$) and to the onset of the near-field regime ($t_{nf}$), see Figure 6. The difference between the two loading times would represent drainage through the structure. By looking at the $t_{ff}$ values in Table 2 we see that, for a given structure, increasing the load or decreasing the viscosity lead to shorter times in the far-field regime, consistent with reaching a fluid film thickness of $h_o$ more quickly with larger applied load or lower viscosity. We also find that increasing the load and decreasing the viscosity leads to shorter time to contact, consistent with contact facilitated by drainage through the structures. If the structured surfaces reach contact while the flat ones do not could also explain why the work of separation is less for the structured surfaces. Finally, we also observe instances where the work of separation for a structured surface has not yet reached a plateau, is still increasing with loading time, but is less than that of a smooth sample (see the drainage region of Figure 6). We suggest that the presence of this region could imply that the drag reduction due to surface structure is more significant in the peeling mode than that in the normal mode, consistent with flow perpendicular or parallel to cylinders.[67, 68]

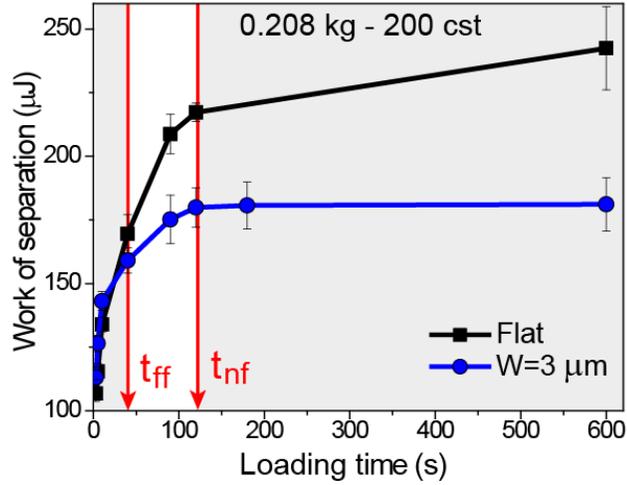

**Figure 6.** Identification of the three regimes from the work of separation as a function of loading time. A schematic of the fluid flow characteristics in each regime is shown on the right. Note that the $t_{nf}$ and $t_{ff}$ are not an interpolated value but determined based on statistical significance.

Based on the scaling argument introduced in Section 2, we predicted that for a given structured surface: 1) $t_{ff} \propto \frac{\mu}{F_N}$, and 2) $t_{ff}$ should be inversely proportional to $h_o^2$. The proportionality between $t_{ff}$ and the ratio $\mu/F_N$ for the two structured surfaces is shown in Figure 7a by using the data for all the loading conditions investigated. For the second prediction we first observe that, as expected, the slope for the data of the W=10μm surface in Figure 7a is less than the one obtained for data coming from the W=3μm surface, consistent with the larger $h_o$ of the W=10μm surface (see Table 1). Finally, we can take the ratio of slopes for the $t_{ff}$ vs $\mu/F_N$ data for the two structured surfaces investigated. Based on Eqn 2, this ratio should be equal to the inverse ratio of $h_o^2$, which is close to what we observe (see Table 3). The range in the calculated values for $h_o$ comes from the fact that it cannot readily be determined for the W=10μm surface because there is not a dominant length scale on the surface features that simplifies the analysis in Refs 34, 35, the same is true for the W=3 μm since the feature sizes are not firmly in the D>>W limit. The relatively good agreement between predictions from Eqn 2 and our measurements indicate that $t_{ff}$ might be a signature for the onset of the drainage through the structures. Finally, based on Eqn 3 we predict that the time to reach boundary contact ($t_{nf}$) should be twice the time spent in the far-field regime, independent of the surface structure. This prediction is confirmed in Figure 7b where $t_{nf}$ is plotted as a function of $t_{ff}$ for the two surfaces investigated and for all the loading conditions. As seen in Figure 7b all the data collapse into a single line of a slope of 1.7. Note here that $h_o$ is calculated based on normal loading (and unloading) and not peeling.

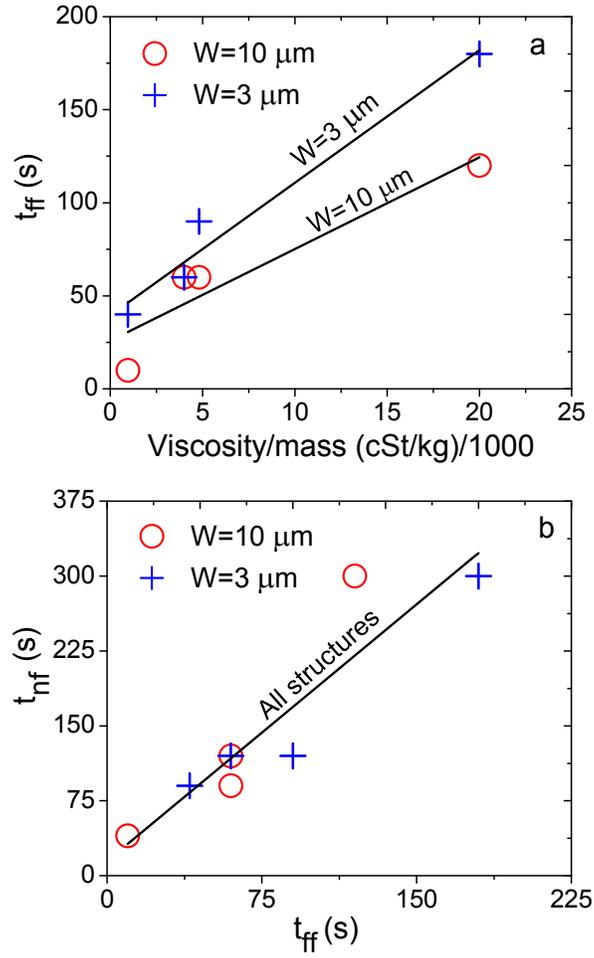

**Figure 7.** (a) Dependence of the far-field limit on the ratio of the ratio of $v/F_N$, based on Eqn 1 we expect the slope to be linear and to be proportional to a length scale unique to each structure geometry. (b) Relationship between the far-field and near-field times, based on Eqn 3 we expect a linear relationship with a slope of 2. Linear least squares fits giving (a) W=3 μm: slope = 0.0071 and $r^2$ = 0.97 and, W=10 μm: slope=0.0049 and $r^2$ = 0.87, and for (b) slope = 1.7 and $r^2$ = 0.85.

**Table 3.** Values extracted from Figure 7. [a] Calculated by using the slopes in Fig. 7a as input in Eqn 2. [b] Values from Table 1. [c] Slope of Fig. 7b. [d] Calculated from Eqn 3.

|  | **Measured** | **Predictions** |
|---|---|---|
| W=10 μm | $h_o$ = 12.0 μm [a] | $h_o$=7.9 μm [b] |
| W=3μm | $h_o$ = 8.1 μm [a] | $h_o$= 2.8-6.1 μm [b] |
| $h_o$(W=3μm)/$h_o$(W=10μm) | 0.69 | 0.34-0.77 |
| $t_{nf}/t_{ff}$ | 1.7 [c] | 2.0 [d] |

We see that estimates for an effective $h_o$ based on our experiments overestimates the predictions of Ref. 34 (see Table 3). The discrepancy could come from the fact that we are using an analysis based on normal motion to describe detachment via small angle peeling measurements. However, in our previous experiments we characterized the normal hydrodynamic forces during approach using the Surface Forces Apparatus. We observed that the onset of reduction of the hydrodynamic force during approach occurred at a separation ($h_c$) that was larger than $h_o$, but we found that $h_c$ was proportional to $h_o$ when comparing different surface structures.[35] Our results here follow the same trend, see the agreement here in the measured and predicted values for $h_o(W=3\mu m)/h_o(W=10\mu m)$ and $t_{ff}/t_{nf}$ in Table 3. Therefore it is more likely that the transition between the different regimes is gradual or occur at separations larger than $h_o$, making $h_o$ an effective measure of the effect of surface structure on viscous forces. Also that these predictions are based on scaling arguments in limiting geometries with many assumptions and as such inherently bring uncertainties.

Implication of our results for the role of viscous contribution on detachment via peeling are the following. 1) The presence of drainage channels reduces the drag upon approach and allows surfaces to make boundary contact faster. 2) The reduction in drag that facilitates approach also allows the surfaces to come apart more easily, as indicated by a decrease in the work of separation with structured surfaces when the loading conditions are kept constant. 3) If the fluid film thickness is too large prior to pull out the surface structures have no influence on the work of separation and behave the same way as smooth surfaces. It is interesting to compare our results to those of Patil et al.[29] where the role of surface structure on the viscous forces measured during peeling was investigated. In their work they observed an increase in the work of separation with structured surfaces. We suspect that we reach different conclusions here because our surface structures are rigid (E=5.6 GPa) while they had a very compliant system therefore surface compliance appears to play a very significant role in modulating the peeling force in viscous environments. Recently Drotlef et al.[39] investigated the effect of surface structure on the adhesion force via normal retraction and on the friction force. In their experiments they did not observe that the fluid viscosity had an effect on the adhesion force (measured by peak force during retraction), and observed that the viscous contribution did not play an important role in their measurements. For the hydrodynamic component, in a system with a normal retraction with a compliant load cell, increasing the viscosity would not change the magnitude of the peak force in pull out measurements if the loading conditions are kept constant. While our experimental system is quite different from the toe pads of tree frogs, our results isolate the contribution of drainage channels to the adhesion force in flooded conditions in a loading scheme that mirrors the mode of tree frog attachment and detachment. In a more realistic system drainage channel could facilitate contact, while in contact conservative forces such as van der Waals interactions would become relevant and surface deformation would play an important role.

## 6. Conclusions

Smooth and structured surfaces were loaded normally towards a flat substrate and peeled off in a viscous Newtonian fluid. The experiments were designed to highlight the interplay between the surface structure and the loading conditions. The effect of structures on the peeling forces was investigated and evidence for three regimes for the work of separation were observed. 1) The far-field regime corresponds to large fluid film thickness prior to pull-out and in this regime there is no effect of surface structure on the work of separation. 2) The drainage through structures regime is very short and corresponds to fluid film thickness that are sufficiently small such that the fluid flows through the structure and, as a result, a decrease in the work of separation compared to flat surfaces is observed. 3) The near-field regime corresponds to interactions between individual pillars and the surface where boundary contact is likely to occur, this regime was characterized by a plateau in the work of separation with loading time. Using simple scaling arguments we found that the boundaries for the different regimes could be related to the surface features via a parameter $h_o$. We also found that the relationship between the loading times for the near-field and far-field was near 2, independent of structure and in agreement with predictions.

## 7. Acknowledgements

The authors are grateful for the support provided by the Office of Naval Research – Young Investigator Award (N000141110629), as well as partial support by the Donors of the American Chemical Society Petroleum Research Fund (Grants No. 51803-ND5). We also thank Arianne Sevilla and Rohini Gupta for help in designing the peeling apparatus and support from the Provost Undergraduate Research Award.

## 8. References cited

1. W. J. P. Barnes, C. Oines and J. M. Smith, *J. Comp. Physiol., A*, 2006, **192**, 1179-1191.
2. T. Endlein, W. J. P. Barnes, D. S. Samuel, N. A. Crawford, A. B. Biaw and U. Grafe, *PLoS One*, 2013, **8**, e73810.
3. T. Endlein, A. Ji, D. Samuel, N. Yao, Z. Wang, W. J. P. Barnes, W. Federle, M. Kappl and Z. Dai, *J. R. Soc., Interface*, 2013, **10**, 20120838.
4. W. Federle, W. J. P. Barnes, W. Baumgartner, P. Drechsler and J. M. Smith, *J. R. Soc., Interface*, 2006, **3**, 689-697.
5. W. J. P. Barnes, *Mrs Bulletin*, 2007, **32**, 479-485.
6. W. J. P. Barnes, P. J. P. Goodwyn, M. Nokhbatolfoghahai and S. N. Gorb, *J. Comp. Physiol., A*, 2011, **197**, 969-978.
7. I. Scholz, W. J. P. Barnes, J. M. Smith and W. Baumgartner, *J. Exp. Biol.*, 2009, **212**, 155-162.

8.  J. M. Smith, W. J. P. Barnes, J. R. Downie and G. D. Ruxton, *J. Comp. Physiol., A*, 2006, **192**, 1193-1204.
9.  J.-H. Dirks and W. Federle, *Soft Matter*, 2011, **7**, 11047-11053.
10. K. Low, Defense Science Research Conference and Expo (DSR), Singapore, 2011.
11. A. Aydin, *Mar. Pet. Geol.*, 2000, **17**, 797-814.
12. B. N. J. Persson, *J. Phys.: Condens. Matter*, 2007, **19**, 376110_376111-376116.
13. B. N. J. Persson, *J. Adhes. Sci. Technol.*, 2007, **21**, 1145-1173.
14. J. F. Vignola and J. A. Judge, *J. Appl. Phys.*, 2008, **104**, 124305_124301-124308.
15. C. Vančura, J. Lichtenberg, A. Hierlemann and F. Josse, *Appl. Phys. Lett.*, 2005, **87**, 162510_162511-162513.
16. T. Lee, E. Charrault and C. Neto, *Adv. Colloid Interface Sci.*, 2014.
17. G. D. Bixler and B. Bhushan, *Soft Matter*, 2013, **9**, 1620-1635.
18. G. Hanna and W. J. P. Barnes, *J. Exp. Biol.*, 1991, **155**, 103-125.
19. C. Y. Hui, N. J. Glassmaker, T. Tang and A. Jagota, *J. R. Soc., Interface*, 2004, **1**, 35-48.
20. J. Y. Chung and M. K. Chaudhury, *J. R. Soc., Interface*, 2005, **2**, 55-61.
21. J. Tamelier, S. Chary and K. L. Turner, *Langmuir*, 2013, **29**, 10881-10890.
22. Y. Tian, N. Pesika, H. B. Zeng, K. Rosenberg, B. X. Zhao, P. McGuiggan, K. Autumn and J. Israelachvili, *Proc. Natl. Acad. Sci. U. S. A.*, 2006, **103**, 19320-19325.
23. B. Zhao, N. Pesika, H. Zeng, Z. Wei, Y. Chen, K. Autumn, K. Turner and J. Israelachvili, *J. Phys. Chem. B*, 2008, **113**, 3615-3621.
24. M.-J. Dalbe, S. Santucci, L. Vanel and P.-P. Cortet, *Soft Matter*, 2014, **10**, 9637-9643.
25. K. Autumn, A. Dittmore, D. Santos, M. Spenko and M. Cutkosky, *J. Exp. Biol.*, 2006, **209**, 3569-3579.
26. A. Ghatak, L. Mahadevan and M. K. Chaudhury, *Langmuir*, 2005, **21**, 1277-1281.
27. A. Ghatak, L. Mahadevan, J. Y. Chung, M. K. Chaudhury and V. Shenoy, *Proc. R. Soc. A*, 2004, **460**, 2725-2735.
28. A. Majumder, A. Sharma and A. Ghatak, *Langmuir*, 2010, **26**, 521-525.
29. S. Patil, R. Mangal, A. Malasi and A. Sharma, *Langmuir*, 2012, **28**, 14784-14791.
30. D. F. Moore, *Wear*, 1965, **8**, 245-263.
31. J. S. Stefan, K., *Sitz. Kais. Akad. Wiss. Math. Natur. Wien.*, 1874, **69**    713-735.
32. O. Reynolds, *Phil. Trans.*, 1886, **177**, 157-234.
33. D. F. Moore, *J. Fluid Mech.*, 1964, **20**, 321-330.
34. B. N. J. Persson, *J. Phys.: Condens. Matter*, 2007, **19**.
35. R. Gupta and J. Frechette, *Langmuir*, 2012, **28**, 14703-14712.
36. O. I. Vinogradova and A. V. Belyaev, *J. Phys.: Condens. Matter*, 2011, **23**, 184104.
37. A. A. Potanin and W. B. Russel, *Phys. Rev. E: Stat., Nonlinear, Soft Matter Phys.*, 1995, **52**, 730-737.
38. C. Wu, H. S. Leese, D. Mattia, R. R. Dagastine, D. Y. Chan and R. F. Tabor, *Langmuir*, 2013, **29**, 8969-8977.
39. D.-M. Drotlef, L. Stepien, M. Kappl, W. J. P. Barnes, H.-J. Butt and A. del Campo, *Adv. Funct. Mater.*, 2013, **23**, 1137-1146.
40. E. Cheung and M. Sitti, *J. Adhes. Sci. Technol.*, 2008, **22**, 569-589.
41. A. E. Kovalev, M. Varenberg and S. N. Gorb, *Soft Matter*, 2012, **8**, 7560-7566.
42. K. Li and S. Cai, *Soft Matter*, 2014, **10**, 8202-8209.
43. R. Gupta and J. Fréchette, *J. Colloid Interface Sci.*, 2013, **412**, 82-88.
44. B. A. Francis and R. G. Horn, *J. Appl. Phys.*, 2001, **89**, 4167-4174.
45. K. Kendall, *J. Phys. D: Appl. Phys.*, 1975, **8**, 1449-1452.
46. J. R. Lister, G. G. Peng and J. A. Neufeld, *Phys. Rev. Lett.*, 2013, **111**, 154501.
47. A. N. Gent and G. R. Hamed, *Rubber Chem. Technol.*, 1979, **52**, 1057-1071.
48. A. T. Al-Halhouli, I. Kampen, T. Krah and S. Büttgenbach, *Microelectron. Eng.*, 2008, **85**, 942-944.

49. J. M. Piau, G. Ravilly and C. Verdier, *J. Polym. Sci. Part B Polym. Phys.*, 2005, **43**, 145-157.
50. S. Patil, A. Malasi, A. Majumder, A. Ghatak and A. Sharma, *Langmuir*, 2011, **28**, 42-46.
51. M. Tirumkudulu, W. B. Russel and T. Huang, *Phys. Fluids*, 2003, **15**, 1588-1605.
52. S. Poivet, F. Nallet, C. Gay and P. Fabre, *Europhys. Lett.*, 2003, **62**, 244.
53. A. Ghatak and M. K. Chaudhury, *J. Adhes.*, 2007, **83**, 679-704.
54. A. J. Crosby, K. R. Shull, H. Lakrout and C. Creton, *J. Appl. Phys.*, 2000, **88**, 2956-2966.
55. N. Canas, M. Kamperman, B. Volker, E. Kroner, R. M. McMeeking and E. Arzt, *Acta Biomater.*, 2012, **8**, 282-288.
56. M. Varenberg and S. N. Gorb, *Adv. Mater.*, 2009, **21**, 483-+.
57. A. Majumder, A. Ghatak and A. Sharma, *Science*, 2007, **318**, 258-261.
58. A. Ghatak, *Phys. Rev. E: Stat., Nonlinear, Soft Matter Phys.*, 2010, **81**.
59. D. Kaelble, *J. Rheol.*, 1959, **3**, 161-180.
60. J. Bikerman and W. Yap, *J. Rheol.*, 1958, **2**, 9-21.
61. D. R. Helsel and R. M. Hirsch, *Statistical methods in water resources*, Elsevier, Amsterdam ; New York, 1992.
62. M. Scaraggi and B. N. J. Persson, *Tribol. Lett.*, 2012, **47**, 409-416.
63. C. Cawthorn and N. Balmforth, *J. Fluid Mech.*, 2010, **646**, 327-338.
64. B. Persson and M. Scaraggi, *J. Phys.: Condens. Matter*, 2009, **21**, 185002.
65. C. S. Davis, D. Martina, C. Creton, A. Lindner and A. J. Crosby, *Langmuir*, 2012, **28**, 14899-14908.
66. T. J. Bowman, G. Drazer and J. Frechette, *Biomicrofluidics*, 2013, **7**, 064111.
67. J. Happel, *AIChE J.*, 1959, **5**, 174-177.
68. A. Sangani and A. Acrivos, *Int. J. Multiphase Flow*, 1982, **8**, 193-206.

# Supplementary Information (ESI)

**Peeling Apparatus.**

The bath is made of acrylic with one optically clear viewing wall (Edmund Optics), and the fluid volume in the bath is approximately 45 mL. Prior to testing, samples are spin coated at 7000 rpm for one minute with the same fluid used in the bath to ensure that fluid is present in the structures and to avoid trapped bubbles in the structures.

**Loading phase**

An aluminum foil boat, visible in Fig. S1, is placed between the weight and the upper surface to be able to remove the weight without pulling the surfaces apart at the end of the loading phase. There is no fluid in the boat, therefore no viscous forces are present when removing the weight. In the absence of the boat the viscous forces between the coverslip and the weight would pull the coverslip away from the bottom surface and increase the fluid film thickness. The aluminum boat also does not prevent the coverslip from bending during peeling phase.

**Peeling phase.**

The rigid contactor pulling on the edge of the coverslip is mounted onto a bending beam load cell (Model LCL-454, Omega Engineering with a DP7600 strain meter, Omega Engineering, 0.4 mN resolution, ~20 readings per second). The bending beam load cell is attached to a vertically translating motorized stage (NSL4 Precision Linear Stage with custom 10:1 planetary gear, Newmark Systems, 0.13 μm resolution). During the peeling phase, a CCD camera (AVT Stingray F-125, binned to 644x300, 70 FPS) was used to take images of the sample as shown in Figure 2. The motor, camera, and data acquisition are all controlled through LabVIEW (National Instruments).

A representative force curve obtained during the peeling phase is shown in Figure S1 along with snapshots of the surfaces taken at different times during the peeling process. The force is plotted as a function of the motor displacement. Based on the pictures, we see that initially the separation between the surfaces appears constant, and the section of the coverslip in the bulk of the fluid up to the contactor begins to bend (Figure S1a) until the force reaches a peak (Figure S1b). Right after the peak, as the contactor continues to move upward, a crack becomes clearly visible and propagates laterally (Figure S1c) as the force decreases abruptly and the contactor tilts upward.

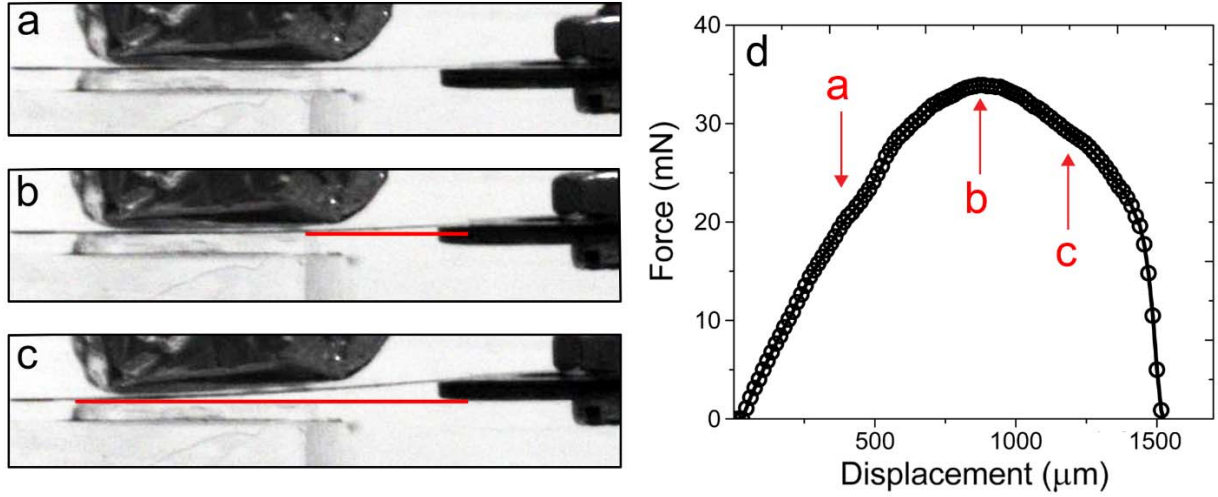

**Figure S1.** (a,b,c) Side view images taken during peeling. The sample is a smooth surface in 1000 cSt silicone oil brought near the lower surface with a 0.05 kg load during 180 seconds. The arrows on the force curve in (d) correspond to the images of (a-c). The red lines in (b,c) are a visual guide outlining the bottom substrate to show bending in the sample, which is slightly visible in (b) and more prominent in (c). The normal force applied by the aluminum foil is negligible during the peeling phase.

**Calculation of the van der Waals interaction between two semi-infinite media**

The non-retarded Hamaker constant[1] was calculated below, assuming a constant value for the dielectric constant.

$$A_{123} \approx \frac{3}{4} kT \left(\frac{\varepsilon_1 - \varepsilon_3}{\varepsilon_1 + \varepsilon_3}\right)\left(\frac{\varepsilon_2 - \varepsilon_3}{\varepsilon_2 + \varepsilon_3}\right) + \frac{3h\nu_e}{8\sqrt{2}} \frac{(n_1^2 - n_3^2)(n_2^2 - n_3^2)}{(n_1^2 + n_3^2)^{\frac{1}{2}}(n_2^2 + n_3^2)^{\frac{1}{2}}((n_1^2 + n_3^2)^{\frac{1}{2}} + (n_2^2 + n_3^2)^{\frac{1}{2}})}$$

Where 1 = CyTop, 2 = Silicone Oil (intervening medium) and 3 = SU-8, k is the Boltzmann constant, T is the temperature (298 K), h is the Planck constant and $\nu_e$ is the main electronic absorption frequency in the UV spectrum. The parameters are listed in Table S1. Based on these value we obtain a Hamaker constant of $A_{CyTop-Silicone\ Oil-SU-8} = -8.0 \times 10^{-23}\ J$.

**Table S1:** Material properties.

| Material | Dielectric Constant | Index of Refraction |
|---|---|---|
| CyTop | 2[2] | 1.34[2] |
| Silicone Oil | 2.74[3] | 1.4[4] |
| SU-8 | 3.2[5] | 1.39[5] |
| $\nu_e$=3E15 s$^{-1}$ | | |

The van der Waal interaction energy (per unit area) of two flat surfaces[1] was calculated using:

$$W = -\frac{A_{CyTop-Silicone\ Oil-SU-8}}{12\pi D^2}$$

At a separation D=2 nm,

$$W = 5.31\ \text{E} - 7\ \frac{J}{m^2}$$

Considering the interacting area between the surface and substrate is a square with a 14 mm length, the van der Waals interaction energy between the two substrates would be 1.04 10$^{-10}$ J, far below the typical work of separation measured in our system.

**References cited**

1. J. N. Israelachvili, *Intermolecular and surface forces: revised third edition*, Academic press, 2011.
2. CYTOP Technical Data, www.bellexinternational.com/products/cytop/pdf/cytop-catalog-p8.pdf, Accessed Dec. 14, 2014.
3. XIAMETER® PMX-200 SILICONE FLUID, www.xiameter.com/EN/Products/Pages/ProductDetail.aspx?pid=01013190&lir=X76, Accessed Dec. 14, 2014.
4. Dielectric Properties of Pure Silicone Fluids, www.clearcoproducts.com/pdf/library/Dielectrical-Properties1.pdf, Accessed Dec. 17, 2014.
5. SU-8 2000 Permanent Epoxy Negative Photoresist, www.microchem.com/pdf/SU-82000DataSheet2025thru2075Ver4.pdf, Accessed Dec. 14, 2014.